\documentclass[aps,prl,twocolumn,groupedaddress,superscriptaddress,showpacs]{revtex4-1}

\usepackage{graphicx}
\usepackage{colortbl}
\usepackage{ulem}

\makeatletter
\def\btt#1{\texttt{\@backslashchar#1}}%
\DeclareRobustCommand\bblash{\btt{\@backslashchar}}%
\makeatother

\begin{document}


\title{Slow steady flow of a skyrmion lattice in a confined geometry probed by resistance narrow-band noise}

\author{Takuro~Sato}
\email{takuro.sato@riken.jp}
\affiliation{RIKEN Center for Emergent Matter Science (CEMS), Wako 351-0198, Japan}

\author{Wataru~Koshibae}
\affiliation{RIKEN Center for Emergent Matter Science (CEMS), Wako 351-0198, Japan}

\author{Akiko~Kikkawa}
\affiliation{RIKEN Center for Emergent Matter Science (CEMS), Wako 351-0198, Japan}

\author{Tomoyuki~Yokouchi}
\affiliation{RIKEN Center for Emergent Matter Science (CEMS), Wako 351-0198, Japan}

\author{Hiroshi~Oike}
\affiliation{RIKEN Center for Emergent Matter Science (CEMS), Wako 351-0198, Japan}
\affiliation{Department of Applied Physics, University of Tokyo, Tokyo 113-8656, Japan}

\author{Yasujiro~Taguchi}
\affiliation{RIKEN Center for Emergent Matter Science (CEMS), Wako 351-0198, Japan}

\author{Naoto~Nagaosa}
\affiliation{RIKEN Center for Emergent Matter Science (CEMS), Wako 351-0198, Japan}
\affiliation{Department of Applied Physics, University of Tokyo, Tokyo 113-8656, Japan}

\author{Yoshinori~Tokura}
\affiliation{RIKEN Center for Emergent Matter Science (CEMS), Wako 351-0198, Japan}
\affiliation{Department of Applied Physics, University of Tokyo, Tokyo 113-8656, Japan}

\author{Fumitaka~Kagawa}
\email{kagawa@ap.t.u-tokyo.ac.jp}
\affiliation{RIKEN Center for Emergent Matter Science (CEMS), Wako 351-0198, Japan}
\affiliation{Department of Applied Physics, University of Tokyo, Tokyo 113-8656, Japan}

\begin{abstract}
Using resistance fluctuation spectroscopy, we observe current-induced narrow-band noise (NBN) in the magnetic skyrmion-lattice phase of micrometer-sized MnSi. 
The NBN appears only when electric-current density exceeds a threshold value, indicating that the current-driven motion of the skyrmion lattice triggers the NBN.
The observed NBN frequency is 10--10$^4$ Hz at $\sim$10$^{9}$ A/m$^{2}$, implying a skyrmion steady flow velocity of 1--100 $\mu$m/s, 3--5 orders of magnitude slower than previously reported. 
The temperature evolution of the NBN frequency suggests that the steady flow entails thermally activated processes, which are most likely due to skyrmion creation and annihilation at the sample edges. 
This scenario is qualitatively supported by our numerical simulations considering boundary effects, which reveals that the edges limit the steady flow of skyrmions, especially at low temperatures.
We discuss a mechanism that dramatically slows the skyrmion steady flow in a microfabricated specimen.
\end{abstract}

\pacs{}

\maketitle

\section{I. INTRODUCTION}
The continuity equation in hydrodynamics describes the flow of conserved particles: $\frac{\partial}{\partial x} \left[  \rho(x, t)v(x, t)  \right] +  \frac{\partial}{\partial t} \left[  \rho(x, t) \right] = 0$ (for a one-dimensional flow), where $\rho(x, t)$ and $v(x, t)$ are the density and velocity of a considered particle, respectively, at position $x$ and time $t$. 
When considering a steady-flow state, the time derivative is zero, and thus, the equation is simplified as: $\frac{\partial}{\partial x} \left[  \tilde{\rho}(x)\tilde{v}(x)  \right] = 0 $ (the tilde denotes a time-averaged value), indicating that $\tilde{\rho}(x)\tilde{v}(x)$ is a constant, $\rho_{0}v_{0}$, independent of $x$. 
This simplest situation can easily be seen, for instance, when electrons steadily flow through a closed-loop electric circuit. 
When electric wires are connected with a conducting material and a d.c.~electric field is applied to the circuit, the resulting electric current is continuous ($\frac{\partial}{\partial x} \left[  \tilde{\rho}(x)e\tilde{v}(x)  \right] = 0 $) although the closed-loop circuit is composed of different types of electric elements. During the circulation of the electric current in the closed-loop circuit, electrons pass through the electrodes (i.e., the boundaries between the electric wires and the material) without being created and annihilated.

An ensamble of magnetic skyrmions \cite{Bogdanov_Sov.Phys.JETP, Bogdanov_J.Magn.Magn.Mater, Muhlbauer_Science, Munzer_Phys.Rev.B, Yu_Nature}, topologically protected particle-like objects, is emerging exotic fluids that flow under an electric current \cite{Jonietz_Science, Schulz_Nat.Phys., Yu_Nat.Commun., Fert_Nat.Nanotech., Sampaio_Nat.Nanotech., Iwasaki_Nat.Commun., Iwasaki_Nat.Nanotech., Woo_Nat.Mat., Jiang_Nat.Phys., Litzius_Nat.Phys.}. 
However, unlike the example of electrons, skyrmions cannot flow through a closed-loop electric circuit that consists of a skyrmion-hosting material and electric-current-carrying metallic wires connected with it, because skyrmions cannot exist in normal nonmagnetic metals, such as copper. Thus, to sustain a skyrmion steady flow, skyrmions need to be created or annihilated at the edges of the material (i.e., the boundaries between the normal metallic wires and the skyrmion-hosting material).
From the point of view of the continuity equation, this situation can be represented by introducing a skyrmion sink/source term $\mathcal{T}_{\rm{sink/source}}(t)$ at the edges, and one can then obtain $\rho_{0}v_{0}=\tilde{\mathcal{T}}_{\rm{sink/source}}$. 
This result implies that a skyrmion steady flow is dictated by a nonlocal interaction between the inside and edges of the specimen, rather than by a local current-skyrmion interaction. 
For instance, when topological skyrmions are so robust that they cannot be either created or annihilated, $\mathcal{T}_{\rm{sink/source}}(t)$ is invariably zero, and hence, $\rho_{0}v_{0}\equiv0$; namely, the skyrmion flow is stopped by the presence of edges. 
Real systems more or less deviate from this extreme situation, thereby allowing for a skyrmion steady flow from one edge to the other \cite{Jonietz_Science, Schulz_Nat.Phys.}. 
Nevertheless, the velocity of this flow can be strongly influenced by the skyrmion creation and/or annihilation rates at the edges, especially in small-sized systems. This issue remains unexplored thus far.

\begin{figure}
\includegraphics[width=8.6cm,clip]{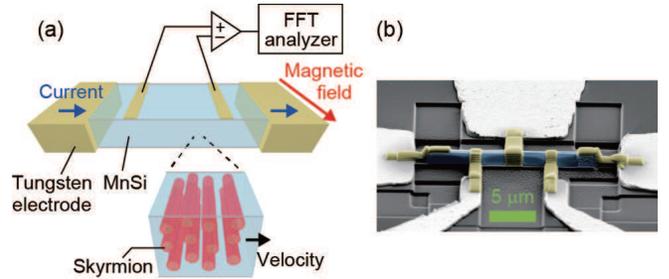}
\caption{(Color online) 
(a) Schematic of the experimental configurations. 
The external magnetic field is applied parallel to the substrate and perpendicular to the long axis of the bar-shaped sample. 
The electric current is applied along the long axis. 
(b) Scanning electron microscope image of the microfabricated MnSi used in this study: 
MnSi crystal (blue), gold electrodes (white), tungsten (yellow) to fix the MnSi and to connect the gold electrodes to MnSi, and a silicon stage (dark gray). 
The scale bar is 5 $\mu$m. 
The center electrode in (b) is not used in the present measurements.
}
\label{Fig1} 
\end{figure}

\begin{figure}
\includegraphics[width=8.6cm,clip]{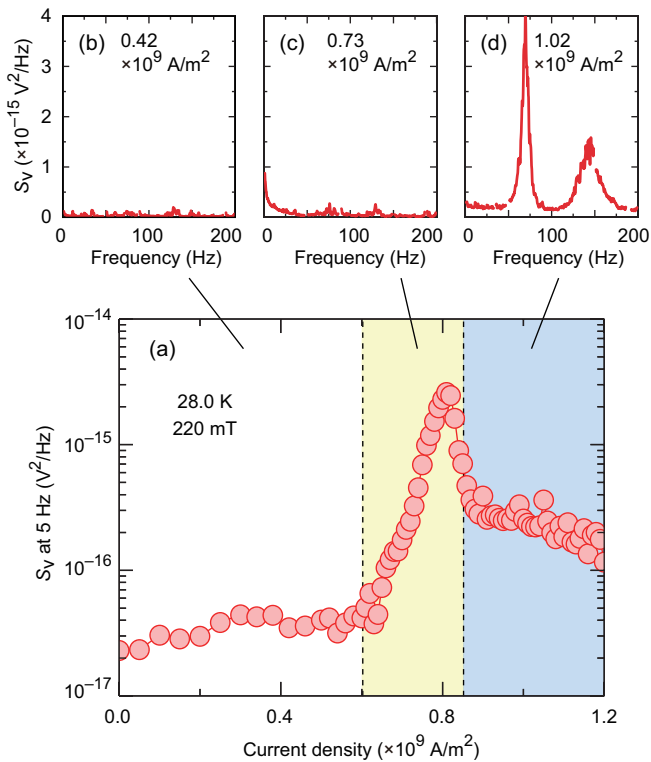}
\caption{(Color online) 
(a) Voltage  power spectral density (PSD) at 5 Hz,  $S_{\rm{V}}$(5 Hz), as a function of applied current density, obtained in the skyrmion lattice phase (28.0 K and 220 mT; for the magnetic phase diagram of the specimen, see Fig.~S1 in Supplemental Material \cite{Supplemental}).
Overall, three dynamical regimes are discernible, as highlighted with different colours (white, yellow, and blue). 
(b--d) The typical PSD in each regime highlighted in (a). 
}
\label{Fig2} 
\end{figure}

In this paper, to address a fundamental question that what impact the sample edges have on the steady flow of topologically protected objects, we investigate the skyrmion steady flow in a confined geometry and scrutinize fluctuations of the longitudinal resistance, $\delta R$, using the setup schematically shown in Fig.~1(a) (for details, see the Method section in Supplemental Material \cite{Supplemental}). 
This approach is underpinned by a recent numerical study \cite{Diaz_Phys.Rev.B}, which reveals that (i) when a skyrmion aggregate flows approximately along the current direction, an intrinsic triangular lattice is dynamically formed; 
(ii) nevertheless, due to the presence of impurities, the velocity vector still contains isotropically fluctuating components, which result in narrow-band noise (NBN) in the velocity power spectral density (PSD) along any direction; and
(iii) most importantly, the mean skyrmion velocity, $|\tilde{v}|$, can be derived from the fundamental frequency of the NBN, as detailed below. 
Such a temporal modulation of the skyrmion velocity,  $\delta v$, may yield resistance fluctuations in a skyrmion-hosting material, and we therefore pursued NBN in an electric field.

\section{II. EXPERIMENTS}
As a target system, we chose MnSi, a material that hosts string-like skyrmions in a triangular-lattice form \cite{Muhlbauer_Science}.
A MnSi single crystal was grown by the Czochralski method. 
To highlight a possible role of edges in skyrmion steady flow, we prepared a sample with a large surface-to-volume ratio ($\sim$$14\times1\times0.8$ $\mu$m$^{3}$) using a focused ion beam (FIB) [Fig.~1(b)]. We found that our microfabricated MnSi specimen exhibits a similar magnetic phase diagram to that of bulk MnSi (see Fig.~S1 in Supplemental Material \cite{Supplemental}). Note that even smaller size of MnSi is known to host a skyrmion lattice phase \cite{Yu_Phys.Rev.B}.

We measured the voltage PSD, $S_{\rm{V}}(f)$,  using the conventional four-terminal d.c.~method as a function of d.c.~electric current. 
A steady current along the long axis of the specimen was supplied by a low-noise voltage source, and the voltage between the voltage-probing electrodes was fed into a spectrum analyzer (Agilent, 35670A) after amplifying it with a low-noise preamplifier (NF Corporation, SA-400F3), as schematically shown in Fig.~1(a).
A large-load resistor was used to eliminate the effect of voltage fluctuations that occur at the contacts of the current leads.

\begin{figure*}
\includegraphics[width=17.8cm,clip]{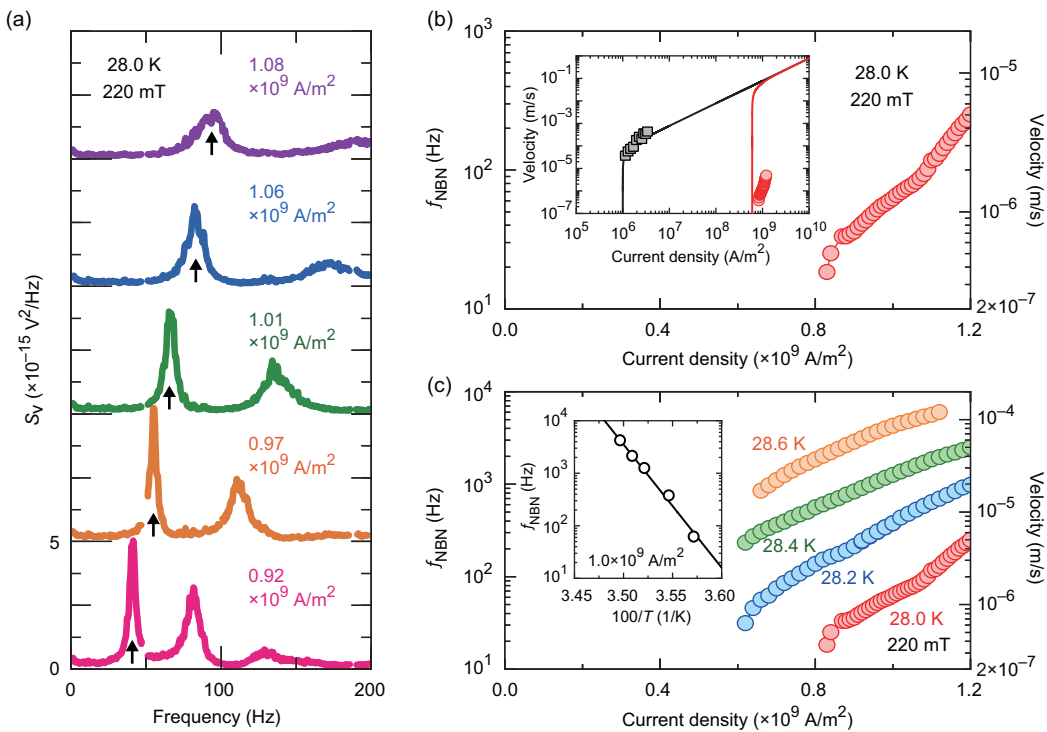}
\caption{(Color online) 
(a) Voltage power spectral density exhibiting NBN for varying current densities in the skyrmion lattice phase. 
Black arrows indicate the fundamental frequency of the NBN, $f_{\rm{NBN}}$, at each current density. 
Each spectrum is vertically offset for clarity. 
(b) Current density dependence of $f_{\rm{NBN}}$ derived from (a). The corresponding velocity of the skyrmion 
steady flow is represented on the right axis, derived from the equation $f_{\rm{NBN}} = |\tilde{v}|/a$ (see the main text). 
Inset: Comparison of the present results (circles) with the experimental results in bulk MnSi (squares \cite{Schulz_Nat.Phys.}) and the theory incorporating the material parameters of MnSi, $v \approx 7.7\times10^{-11}(j^{2} - j_{\rm{c}}^{2})^{0.5}$ (solid lines \cite{Iwasaki_Nat.Commun.}: 
the units of $v$ and $j$ are m/s and A/m$^{2}$, respectively). 
In the fitting, the $j_{\rm{c}}$ values are taken from the corresponding experiment: $j_{\rm{c}} = 1\times10^{6}$ (A/m$^{2}$) for the black line \cite{Schulz_Nat.Phys.} and $j_{\rm{c}}= 0.6\times10^9$ (A/m$^{2}$) for the red line from this study. 
(c) $f_{\rm{NBN}}$-current profiles at different temperatures. 
Inset: Arrhenius plot of $f_{\rm{NBN}}$ at a certain current density, 1$\times$10$^{9}$ A/m$^{2}$. 
The activation energy, $\approx$5400 K, only weakly depends on the choice of the current density; see Fig.~S6 in Supplemental Material \cite{Supplemental}. 
}
\label{Fig3}
\end{figure*}

\section{III. RESULTS AND DISCUSSIONS}
Figure 2(a) displays systematic variations of $S_{\rm{V}}$ at a low frequency, 5 Hz, in the skyrmion lattice phase (The selection of a frequency does not affect the following arguments; for details, see Fig. S4 in Supplemental Material \cite{Supplemental}). 
As highlighted with different colors, the overall behavior can be categorized into three dynamical regimes, and the typical spectrum in each regime is displayed in Figs.~2(b)--2(d). 
At low current densities, no notable evolution is observed in the spectral shape and intensity [Fig.~2(b)], indicating that all skyrmions remain stationary. 
However, when the current density reaches a critical value, $j_{\rm{c}}$ ($\approx$$0.6\times10^9$ A/m$^{2}$), a $1/f^{\alpha}$ spectrum with $1<\alpha<2$ appears [Fig.~2(c)], signaling the onset of skyrmion motion: we call this featureless spectrum broad-band noise (BBN). 
As the current density further increases, we find that the BBN eventually turns into NBN accompanied by a higher harmonic [Fig.~2(d)], a key finding of this study.
Such successive dynamical transitions are observed only in the skyrmion lattice phase and when the electric current is applied normal to the magnetic field (see Figs.~S1--S3 and S5 in Supplemental Material \cite{Supplemental}), leading us to conclude that the observations reflect current-induced dynamics of skyrmions.
Our observations---namely, the emergence of BBN, followed by the formation of NBN, whose fundamental frequency increases with the applied current---reproduce the main characteristics of those found for the current-induced motions of charge/spin density waves \cite{Fleming_Phys.Rev.Lett., Hundley_Phys.Rev.B, Sekine_Phys.Rev.B} and superconducting-vortex lattices \cite{Togawa_Phys.Rev.Lett.}, and they are also in good agreement with the results of numerical simulations for skyrmions \cite{Diaz_Phys.Rev.B, Reichhardt_Phys.Rev.Lett., Reichhardt_New.J.Phys., Koshibae_Sci.Rep.}. 
On the basis of these analogies, we may reasonably attribute the observed BBN and NBN to current-induced spatiotemporally incoherent and coherent steady flows of the skyrmions, respectively.

To gain more insight into the NBN, we investigated the detailed current dependence.
The fundamental frequency of the NBN, $f_{\rm{NBN}}$, continuously shifts toward higher frequencies as the current increases [Figs.~3(a) and 3(b)], giving invaluable information about the mean skyrmion velocity, $|\tilde{v}|$. 
In determining $|\tilde{v}|$ from $f_{\rm{NBN}}$, we refer to a standard model, $f_{\rm{NBN}}=|\tilde{v}|/a$, where $a$ is the characteristic periodicity of a long-range-ordered system under consideration. 
This equation has been experimentally applied to various systems exhibiting NBN, such as charge/spin density waves \cite{Fleming_Phys.Rev.Lett., Hundley_Phys.Rev.B, Sekine_Phys.Rev.B} and superconducting-vortex lattices \cite{Togawa_Phys.Rev.Lett.}; in addition, this equation is also verified in numerical simulations for a skyrmion-lattice motion under d.c.~driving force \cite{Diaz_Phys.Rev.B}.
Depending on microscopic dynamics of the skyrmion lattice, one may consider a different length scale, such as the sample length \cite{Tsuboi_Phys.Rev.Lett.}. However, given no experimental/theoretical basis supporting such a scenario at present, here we took 20 nm (the skyrmion lattice constant) as $a$ \cite{Muhlbauer_Science} and derived $|\tilde{v}|$, as displayed on the right axis of Fig.~3(b). 
The derived velocity, 0.4--4\ $\mu$m/s, for this microfabricated sample is substantially low compared with the experimental results for bulk MnSi \cite{Schulz_Nat.Phys.} and numerical results for a periodic-boundary system \cite{Iwasaki_Nat.Commun.}, although the two reference data agree with each other as long as an appropriate $j_{\rm{c}}$ value is considered [Fig.~3(b), inset]. 
This large discrepancy, 5--6 orders of magnitude, suggests that the microfabricated sample has a unique mechanism significantly slowing the skyrmion steady flow.

A plausible explanation is that the creation and/or annihilation rates of skyrmions at the edges (namely, $\tilde{\mathcal{T}}_{\rm{sink/source}}$) are particularly low in the microfabricated sample and thus limit the steady flow.
 In this context, it should first be noted that the creation or annihilation of a topologically protected object generally requires a thermally activated, discontinuous process. 
Conversely, if the skyrmion steady-flow velocity involves skyrmion creation and annihilation processes, it would exhibit substantially faster dynamics at higher temperatures.
In this context, we investigated the temperature evolution of the $j$--$f_{\rm{NBN}}$ profiles, and the results are diplayed in Fig.~3(c).
In this narrow temperature range (28.0--28.6 K), $f_{\rm{NBN}}$ clearly increases by more than one order of magnitude, consistent with the expectation. 
The application of the Arrhenius plot, on trial, shows a large activation barrier of $\approx$5400 K [Fig.~3(c), inset; see also Fig.~S6 in Supplemental Material \cite{Supplemental}], featuring the dramatic temperature dependence of the skyrmion velocity.
Importantly, such thermally activated behavior is absent in bulk MnSi \cite{Schulz_Nat.Phys.}, which is reasonably understood if the edges play only a minor role in the bulk sample.

We thus return our attention to the central question of why the skyrmion creation and/or annihilation rates can be so low in the microfabricated sample. 
This issue appears to be explained by considering thermally populated ``emergent magnetic monopoles'', magnetic singularities that can initiate the destruction of the skyrmion string \cite{Milde_Science, Schutte_Phys.Rev.B}. 
Compared with a bulk sample, the skyrmion string length is much shorter in the present specimen, only $\approx$1 $\mu$m. 
This aspect makes the thermal population of emergent magnetic monopoles in each skyrmion string much lower \cite{Kagawa_Nat.Commun., Oike_Phys.Rev.B}, leading to more robust topological protection.
 In fact, compared with bulk MnSi \cite{Oike_Nat.Phys.}, the skyrmion phase in the microfabricated sample can be thermally quenched much more easily (see Fig.~S7 in Supplemental Material \cite{Supplemental}), corroborating the low annihilation rate of the skyrmions. Thus, by considering the continuity equation for a steady-flow state, $\rho_{0}v_{0}=\tilde{\mathcal{T}}_{\rm{sink/source}}$, one can expect a slow steady flow of skymions.

\begin{figure}
\includegraphics[width=8.6cm]{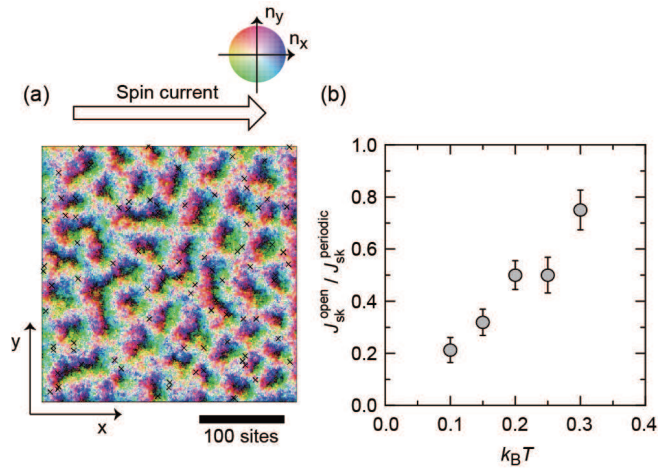}
\caption{(Color online) 
(a) Snapshot of the skyrmions in a steady-flow state under the application of the spin current in the open-boundary system at $k_{\rm{B}}T=0.3$. The simulated lattice consists of 300$\times$300 magnetic moments.
The magnitude and direction of the applied spin current are $v_{\rm{s}} = 0.006$ (corresponding to an electric current of $\sim$6$\times$10$^{10}$ A/m$^{2}$) and the $+x$ direction, respectively. 
The color wheel specifies the $x$-$y$ plane magnetization direction. 
The brightness of the color represents the $z$ component of the magnetization; that is, the local magnetizations pointing in the $z$ direction are displayed as white. 
Crosses indicate the position of impurities. 
(b) Temperature profile of the ratio between $J_{\rm{sk}}^{\rm{open}}$ and $J_{\rm{sk}}^{\rm{periodic}}$, which represent the skyrmion flux in a steady state in the open- and periodic-boundary systems, respectively.
}
\label{Fig4} 
\end{figure}

Finally, to see whether a minimal model can reproduce the interplay between the sample edges and the skyrmion steady flow, we numerically investigated the current-induced dynamics of the skyrmion lattice at finite temperatures in open- and periodic-boundary two-dimensional (2D) systems (see the Method section in Supplemental Material \cite{Supplemental}).
We found that in the open-boundary system, thermal agitation allows finite nucleation and annihilation rates of skyrmions at the edges, eventually resulting in a steady-flow state, in which the two rates are balanced. We note that this is not the case for zero temperature: At a small spin current density, $v_{\rm{s}} = 0.001$ (corresponding to an electric current of $\sim$1$\times$10$^{10}$ A/m$^{2}$), for instance, a skyrmion steady flow is realized in the periodic boundary system, but the skyrmion lattice settles into a stationary state, $|\tilde{v}|= 0$, in the open-boundary system (Fig.~S8 in Supplemental Material \cite{Supplemental}).
Figure 4(a) shows a typical snapshot in the steady-flow state at $k_{\rm{B}}T=0.3$ and $v_{\rm{s}} = 0.006$ (see also Supplemental Movie S1 \cite{Supplemental}).
From its time evolution, we derived the number of skyrmions that pass through one edge per unit time (namely, skyrmion flux, $J_{\rm{sk}}^{\rm{open}}$) in the steady-flow state and compared it with the value in the periodic-boundary system, $J_{\rm{sk}}^{\rm{periodic}}$. 
The ratio, $J_{\rm{sk}}^{\rm{open}}/J_{\rm{sk}}^{\rm{periodic}}$, is found to be invariably less than unity, and moreover, decrease as the temperature is lowered [Fig.~4(b)]. Thus, the simulation results demonstrate that even in the minimal model, the skyrmion steady flow is significantly affected by the skyrmion nucleation/annihilation rates at the edges, consistent with the experiments.
Nevertheless, it is to be noted that the agreement between the experiments and simulations remains only qualitative: The ratio obtained in the simulation is on the order of 0.1 even at the lowest temperature, $k_{\rm{B}}T=0.1$, and it is not as pronounced as $\approx$$10^{-4}$--$10^{-5}$ as observed in our micrometer-sized specimen. 
We tentatively attribute this discrepancy to a difference in the dimension between the numerical simulations (2D) and experiments (3D), and the quantitative agreement remains to be solved.

We expect that similar edge effects can be involved in the steady flow of superconducting vortices because superconducting vortices are also topological objects. 
In fact, the edges of a superconductor are known to work as a barrier when vortex lines are entering and exiting the considered system in response to changes in the external magnetic field \cite{Bean_Phys.Rev.Lett., Schweigert_Phys.Rev.Lett.}. 
However, to our knowledge, no experimental report on the edge-dominated, sluggish steady flow of superconducting vortices exists, implying that in most experiments, the superconducting specimen is not in the clean limit and that the edge effect is less important than the pinning by impurities.

\section{IV. CONCLUSIONS}
We have performed resistance fluctuation spectroscopy on micrometer-sized MnSi and observed current-induced narrow-band noise in the magnetic skyrmion-lattice phase. 
This observation is ascribed to the steady-flow motion of the skyrmion lattice under an electric current. 
The steady flow velocity estimated from the NBN frequency is anomalously slow, 1--100 $\mu$m/s at $\sim$10$^9$ A/m$^2$, and exhibits thermally activated behavior, markedly different from the previous experimental results on a bulk sample and numerical results on a periodic-boundary system. 
Moreover, our numerical simulations on an open-boundary system reveal that the edges limit the skyrmion motion. 
Thus, our findings suggest a vital role of the edges in skyrmion steady flow, especially in a microstructure.

 T. S. and F. K. thank S. Kaneko, K. Ienaga, S. Okuma, M. Ueda, T. Tamegai and S. Otabe for their valuable discussions. 
This work was partially supported by JSPS KAKENHI (Grant No.~18H05225) and JST CREST Grant Number JPMJCR1874, Japan.

\end{document}